\documentclass[%
 reprint,
 amsmath,amssymb,
 aps,
]{revtex4-2}
\usepackage{xcolor}

\usepackage{graphicx}
\usepackage{dcolumn}
\usepackage{bm}
\usepackage{amsmath}
\usepackage{siunitx}
\usepackage{mathtools}

\begin{document}

\preprint{APS/123-QED}


\title{A biophysical threshold for biofilm formation}

\author{Jenna A. Ott$^{1}$, Selena Chiu$^{1}$, Daniel B. Amchin$^{1}$,  \\
 Tapomoy Bhattacharjee$^{2\dagger}$, and Sujit S. Datta$^{1}$
}
\email{Correspondence to ssdatta@princeton.edu.\\
$^{\dagger}$Current affiliation: National Centre for Biological Sciences, Tata Institute of Fundamental Research, Bellary Road, Bangalore 560065, Karnataka, India.}
\affiliation{%
 $^{1}$Department of Chemical and Biological Engineering, Princeton University, Princeton NJ 08544\\
 $^{2}$Andlinger Center for Energy and the Environment, Princeton University, Princeton NJ 08544}

\date{\today}

\begin{abstract}
Bacteria are ubiquitous in our daily lives, either as motile planktonic cells or as immobilized surface-attached biofilms. These different phenotypic states play key roles in agriculture, environment, industry, and medicine; hence, it is critically important to be able to predict the conditions under which bacteria transition from one state to the other. Unfortunately, these transitions depend on a dizzyingly complex array of factors that are determined by the intrinsic properties of the individual cells as well as those of their surrounding environments, and are thus challenging to describe. To address this issue, here, we develop a generally-applicable biophysical model of the interplay between motility-mediated dispersal and biofilm formation under positive quorum sensing control. Using this model, we establish a universal rule predicting how the onset and extent of biofilm formation depend collectively on cell concentration and motility, nutrient diffusion and consumption, chemotactic sensing, and autoinducer production. Our work thus provides a key step toward quantitatively predicting and controlling biofilm formation in diverse and complex settings.

\end{abstract}

\maketitle

Dating back to their discovery by van Leeuwenhoek over three centuries ago, it has been known that bacteria typically exist in one of two phenotypic states: either as motile, planktonic cells that self-propel using e.g., flagella or pili (``animalcules... moving among one another'' \cite{van1665observations}), or as immobilized, surface-attached biofilms (``little white matter... in the scurf of the teeth'' \cite{vanscurf}). These different states have critical functional implications for processes in agriculture, environment, industry, and medicine. For example, motility-mediated dispersal of planktonic cells enables populations to escape from harmful conditions and colonize new terrain \cite{adler1966science,adler1966effect,saragosti2011directional,fu2018spatial,cremer2019chemotaxis,bhattacharjee2020waves}, underlying infection progression, drug delivery to hard-to-reach spots in the body, food spoilage, interactions with plant roots in agriculture, and bioremediation of environmental contaminants \cite{balzan,chaban,dattapnas,harman,ribet,siitonen,lux,oneil,Gill1977,Shirai2017,thornlow,toley,dechesne,souza,turnbull,watt,babalola,Adadevoh2016,Adadevoh2018,ford07,wang08,reddy,martinez2021active}. In addition, the formation of immobilized biofilms can initiate antibiotic-resistant infections, foul biomedical devices and industrial equipment, or conversely, help sequester and remove contaminants in dirty water \cite{davey2000microbial,hall2004bacterial,mah2003genetic,o2005biofilms,fux2005survival,nicolella2000wastewater,donlan2002biofilms,davies1998involvement}. Hence, extensive research has focused on understanding bacterial behavior in either the planktonic or biofilm state.

For example, studies of planktonic cells have provided important insights into bacterial motility---which can be either undirected \cite{berg2018random,bergecoli,bhattacharjee2019bacterial,bhattacharjee2019confinement} or directed in response to e.g., a chemical gradient via chemotaxis \cite{adler1966effect,adler1966science,saragosti2011directional,fu2018spatial,cremer2019chemotaxis,bhattacharjee2020waves,keller1971traveling,odell1976traveling,keller1975necessary,lauffenburger1991quantitative,seyrich2019traveling,croze2011migration,amchin2021influence}. These processes are now known to be regulated not just by intrinsic cellular properties, such as swimming kinematics and the amplitude and frequency of cell body reorientations, but also by the properties of their environment, such as cellular concentration, chemical/nutrient conditions, and confinement by surrounding obstacles \cite{berg2018random,bergecoli,bhattacharjee2019bacterial,bhattacharjee2019confinement,adler1966effect,adler1966science,saragosti2011directional,fu2018spatial,cremer2019chemotaxis,bhattacharjee2020waves,keller1971traveling,odell1976traveling,keller1975necessary,lauffenburger1991quantitative,seyrich2019traveling,croze2011migration,amchin2021influence}. Thus, the manner in which planktonic bacteria disperse can strongly vary between different species and environmental conditions. 

Similarly, studies of biofilms under defined conditions have also provided key insights---such as by revealing the pivotal role of intercellular chemical signaling in biofilm formation \cite{nadell2008evolution,bassler2006bacterially,davey2000microbial,hall2004bacterial}. In this process, termed quorum sensing, individual cells produce, secrete, and sense freely-diffusible autoinducer molecules, thereby enabling different bacteria to coordinate their behavior \cite{davies1998involvement,sakuragi2007quorum,bassler2006bacterially,miller2001quorum,herzberg2006ydgg,laganenka2018autoinducer,mclean1997evidence,paul2009application}. For example, in many cases, quorum sensing positively controls biofilm formation \cite{herzberg2006ydgg,laganenka2018autoinducer,davies1998involvement,sakuragi2007quorum,mclean1997evidence,cai2016singly,chen2011motility}: autoinducer accumulation above a threshold concentration upregulates the expression of genes involved in biofilm formation, ultimately driving a transition from the planktonic to the biofilm state \cite{nadell2008evolution}. Again, however, the cellular factors that control this transition, such as the autoinducer production rate, diffusivity, and threshold concentration, can strongly vary between different species and environmental conditions.

Because planktonic dispersal and biofilm formation both depend on a dizzyingly complex array of factors, these distinct processes are typically studied in isolation. Thus, while each is well understood on its own, quantitative prediction of the conditions under which a population of planktonic bacteria transitions to the biofilm state---or instead, continues to disperse away and remains in the planktonic state---remains challenging. Here, we address this challenge by developing a mathematical model that describes the essential features of motility-mediated dispersal of planktonic cells and autoinducer-mediated biofilm formation together. Using numerical simulations of this model, we systematically examine the influence of cellular concentration, motility, and chemotactic sensing; nutrient availability, diffusion, and consumption; and autoinducer production, diffusion, and accumulation on biofilm formation. Guided by these results, we establish a universal biophysical threshold that unifies the influence of all these factors in predicting the onset and extent of biofilm formation across different species and environmental conditions. Our work therefore provides a theoretical foundation for the prediction and control of biofilm formation in diverse and complex settings, and yields new quantitative predictions to guide future experiments.\\

\begin{figure*}
\includegraphics[width=17.8cm]{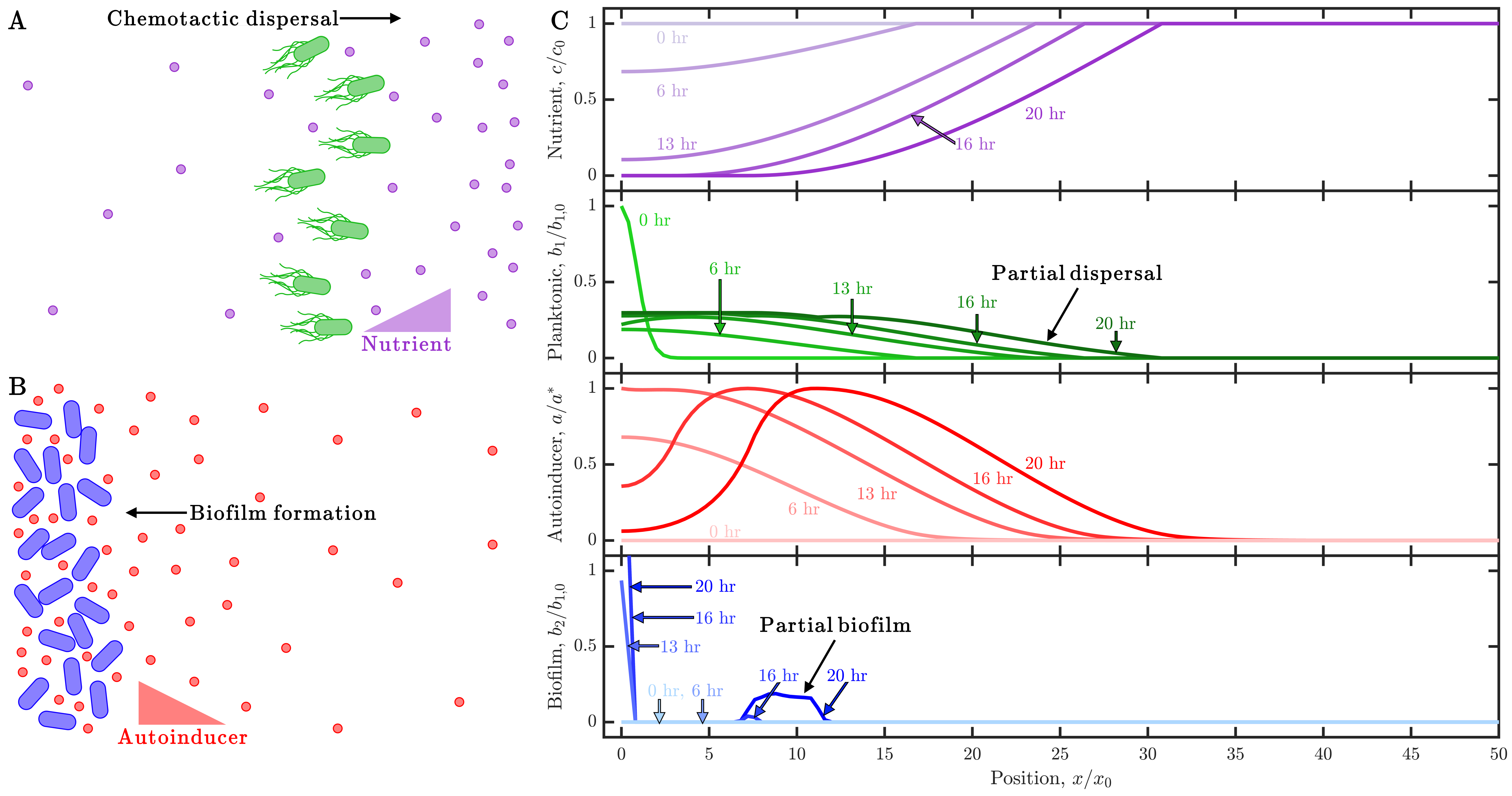}
\caption{\label{fig01} \textbf{Competition between motility-mediated dispersal and autoinducer-mediated biofilm formation.} \textbf{(A)} Schematic of chemotactic dispersal: planktonic bacteria (green) consume nutrient (purple) and establish a local gradient that they, in turn, direct their motion in response to. \textbf{(B)} Schematic of positive quorum sensing-controlled biofilm formation: accumulation of produced autoinducer (red) above a threshold concentration causes cells to transition to the biofilm state (blue). \textbf{(C)} Results of an example simulation of Eqs.~\ref{eq:b1}-\ref{eq:ai} showing the dynamics of the nutrient, planktonic cells, autoinducer, and biofilm cells from top to bottom, quantified by the normalized concentrations $c/c_0$, $b_{1}/b_{1,0}$, $a/a^{*}$, $b_{2}/b_{1,0}$, respectively; $c_0$, $b_{1,0}$, and $a^{*}$ represent the initial nutrient concentration, initial bacterial concentration, and autoinducer threshold for biofilm formation, respectively. The position coordinate is represented by the normalized position $x/x_0$, where $x_0$ is the width of the initial cellular inoculum. Different shades indicate different time points as listed. The inoculum initially centered about the origin consumes nutrient (purple), establishing a gradient that drives outward dispersal by chemotaxis (outward moving green curves); the cells also produce autoinducer (red) concomitantly. At $t\approx13$ \si{\hour}, sufficient autoinducer has been produced to trigger biofilm formation at the origin; at even longer times ($t\gtrsim16$ \si{\hour}), nutrient depletion limits autoinducer production at this position. However, accumulation of autoinducer by the dispersing planktonic cells triggers partial biofilm formation at $x/x_0 \approx 4$ as well. This competition between dispersal and biofilm formation leads to a final biofilm fraction of $f = 21\%$ at the final time of $t=20$ \si{\hour}. An animated
form of this Figure is shown in Movie S1. The values of the simulation parameters are given in Table S2.}
\end{figure*}

\section*{Results}

\subsection*{Development of the governing equations}

As an illustrative example, and to connect our model to recent experiments of bacterial dispersal \cite{bhattacharjee2020waves}, we consider a rectilinear geometry with a starting inoculum of planktonic cells at a maximal concentration $b_{1,0}$ and of width $x_0$. In general, the continuum variable $b(x,t)$ describes the number concentration of bacteria, where $x$ is the position coordinate and $t$ is time, and the subscripts $\{1,2\}$ represent planktonic or biofilm-associated cells, respectively. Following previous work \cite{lauffenburger1991quantitative,keller1971traveling,adler1966science,croze2011migration,fu2018spatial,bhattacharjee2020waves}, we consider a sole diffusible nutrient that also acts as the chemoattractant, with a number concentration represented by the continuum variable $c(x,t)$ with diffusivity $D_{c}$. Initially, nutrient is replete throughout the system at a constant concentration $c_0$. The bacteria then consume the nutrient at a rate $b_{1}\kappa_1 g(c)$, where $\kappa_1$ is the maximum consumption rate per cell and the Michaelis-Menten function $g(c)\equiv \frac{{c}}{{c + c_{\rm{char}}}}$ quantifies the nutrient dependence of consumption relative to the characteristic concentration $c_{\rm{char}}$ \cite{croze2011migration,monod1949growth,cremer2019chemotaxis,woodward1995spatio,shehata1971effect,Schellenberg1977,cremer2016effect}. 

As time progresses, the bacteria thereby establish a local nutrient gradient that they respond to via chemotaxis (Fig.~\ref{fig01}A). In particular, planktonic cells disperse through two processes: undirected active diffusion with a diffusivity $D_{1}$ \cite{berg2018random}, and directed chemotaxis with a drift velocity $\vec{v}_{c}\equiv\chi_{1} \nabla \log\left(\frac{1+c/c_{-}}{1+c/c_{+}}\right)$ that quantifies the ability of the bacteria to sense and respond to the local nutrient gradient  \cite{keller1970initiation,keller1971traveling,odell1976traveling,keller1975necessary} with characteristic bounds $c_{-}$ and $c_{+}$  \cite{cremer2019chemotaxis,sourjik2012responding,shimizu2010modular,tu2008modeling,Kalinin2009,shoval2010fold,lazova2011response,celani2011molecular,fu2018spatial,dufour,yang2015relation} and a chemotactic coefficient $\chi_{1}$. The planktonic cells also proliferate at a rate $b\gamma_{1} g(c)$, where $\gamma_{1}$ is the maximal proliferation rate per cell. Finally, as the planktonic bacteria consume nutrients, they produce and secrete a diffusible autoinducer, with a number concentration represented by $a(x,t)$ and with diffusivity $D_{a}$, at a maximal rate $k_{1}$ per cell. Motivated by some previous work \cite{hense2012spatial,hense2015core,kirisits2007influence,bollinger2001gene,duan2007environmental,mellbye2014physiological,de2001quorum,perez2010heterogeneous}, we take this process (hereafter referred to as ``production'' for brevity) to also be nutrient-dependent via the same Michaelis-Menten function $g(c)$ for the results presented in the main text, but we also consider the alternate case of ``protected'' nutrient-independent production in the SI (Fig. S1). Following previous work \cite{koerber2002mathematical,ward2001mathematical,ward2003early}, we also model natural degradation of autoinducer as a first-order process with a rate constant $\lambda$.

As autoinducer is produced, it binds to receptors on the surfaces of the planktonic cells with a second-order rate constant $\alpha$, as established previously \cite{koerber2002mathematical,ward2001mathematical,ward2003early}. Motivated by experiments on diverse bacteria, including the prominent and well-studied species \textit{Escherichia coli}, \textit{Pseudomonas putida}, and \textit{Pseudomonas aeruginosa} \cite{davies1998involvement,sakuragi2007quorum,bassler2006bacterially,miller2001quorum,herzberg2006ydgg,laganenka2018autoinducer,mclean1997evidence,paul2009application,cai2016singly,chen2011motility}, we assume that planktonic cells transition to the biofilm state when the local autoinducer concentration exceeds a threshold value $a^{*}$ at a rate $\tau^{-1}$ (Fig.~\ref{fig01}B). Because our focus is on this transition, we assume that it is irreversible, and that cells in the biofilm lose motility. However, they still continue to consume nutrient, proliferate, and produce autoinducer with maximal rates $\kappa_2$, $\gamma_{2}$, and $k_2$ per cell, respectively; additional behaviors such as subsequent production of extracellular polymeric substances or transitioning back to the planktonic state can be incorporated as future extensions to this model.

Hence, while planktonic cells can disperse via active diffusion and chemotaxis, their dispersal is hindered---and biofilm formation is instead promoted---when autoinducer accumulates sufficiently, as schematized in Figs.~\ref{fig01}A--B. The central goal of this paper is to examine the processes underlying this competition between dispersal and biofilm formation. Our model is thus summarized as:
\begin{eqnarray}
{\rm{Planktonic:}}~\frac{\partial b_1}{\partial t} &=& \underbrace{D_{1} \nabla^2 b_1 - \nabla \cdot \left( b_1 \vec{v}_{c} \right)}_{\rm{Motility}} + \underbrace{b_1 \gamma_{1} g(c)}_{\rm{Proliferation}}\nonumber
\\
 &&- \underbrace{b_1{\tau}^{-1}\mathcal{H}\left(a - a^{*}\right)}_{\rm{Biofilm~formation}} 
\label{eq:b1}
\\
{\rm{Biofilm:}}~\frac{\partial b_2}{\partial t} &=& \underbrace{b_2 \gamma_{2} g(c)}_{\rm{Proliferation}}
 + \underbrace{b_1{\tau}^{-1}\mathcal{H}\left(a - a^{*}\right)}_{\rm{Biofilm~formation}}
\label{eq:b2}
\\
{\rm{Nutrient:}}~\frac{\partial c}{\partial t} &=& \underbrace{D_{c} \nabla^2 c}_{\rm{Diffusion}} - \underbrace{\left(b_1 \kappa_1+b_2 \kappa_2 \right) g(c)}_{\rm{Consumption}}
\label{eq:c}
\\
{\rm{Autoinducer:}}~\frac{\partial a}{\partial t} &=& \underbrace{D_{a} \nabla^2 a}_{\rm{Diffusion}} + \underbrace{\left(b_1 k_1 + b_2 k_2\right) g(c)}_{\rm{Production}} \nonumber
\\
&&- \underbrace{a\left(\lambda + \alpha b_1 \right)}_{\rm{Loss}}
\label{eq:ai}
\end{eqnarray}
where $\mathcal{H}$ is the Heaviside step function describing the transition from the planktonic to biofilm state. To explore the competition between motility-mediated dispersal and autoinducer-mediated biofilm formation, we then numerically solve this system of coupled equations using values of all parameters---which are either intrinsic descriptors of cellular physiology or are solely/additionally influenced by the local environment---that are derived from experiments (Table S1). Further details are provided in the \textit{Materials and Methods}.

\begin{figure*}
\centering
\includegraphics[width=11.4cm]{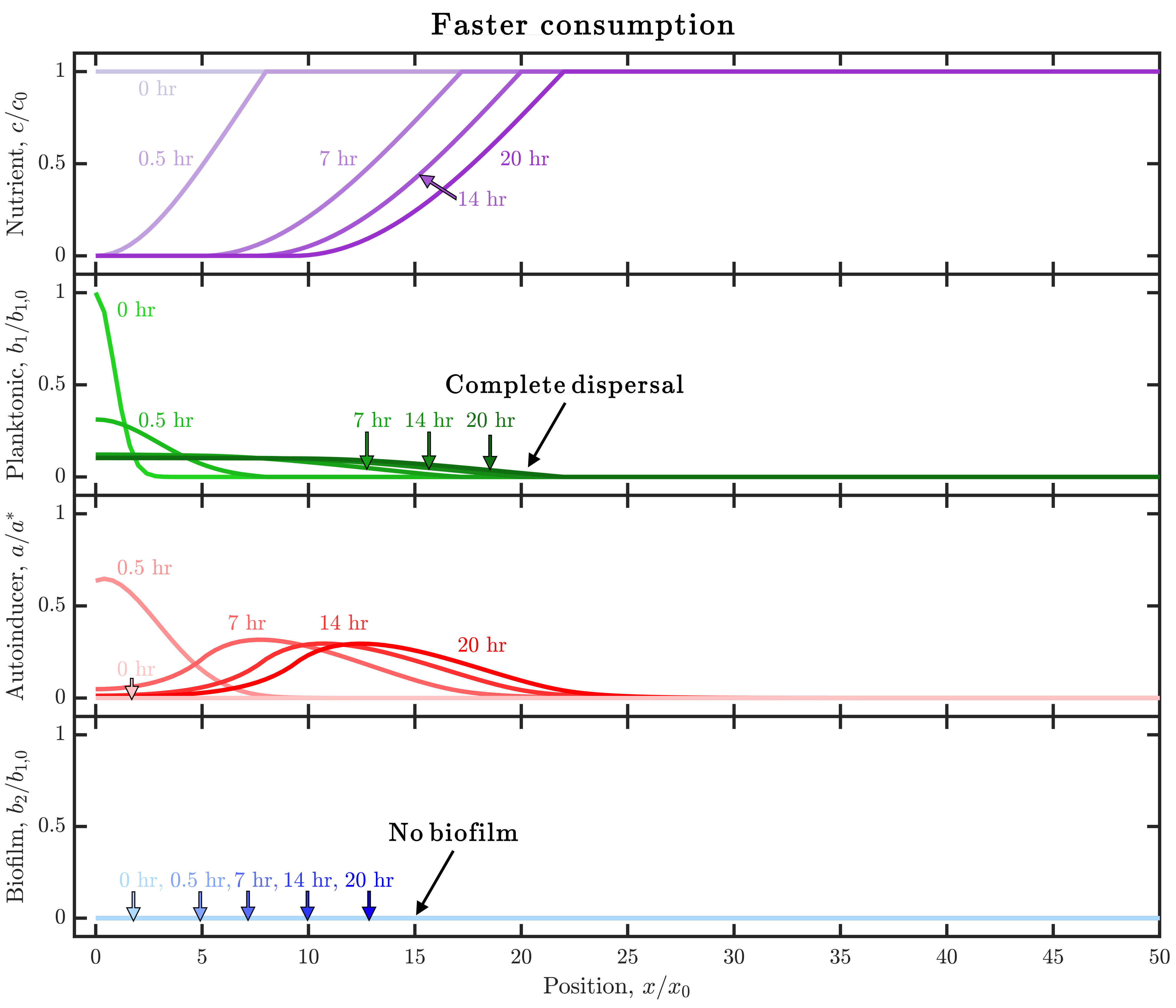}
\caption{\textbf{Faster nutrient consumption limits autoinducer production, leading to complete dispersal.} Results of the same simulation as in Fig.~\ref{fig01}C, but for planktonic cells with faster nutrient consumption (larger $\kappa_1$). Panels and colors show the same quantities as in Fig.~\ref{fig01}C. The inoculum initially centered about the origin consumes nutrient (purple), establishing a gradient that drives outward dispersal by chemotaxis (outward moving green curves); the cells also produce autoinducer (red) concomitantly. However, nutrient is depleted at this position more rapidly, limiting autoinducer production; as a result, the population continues to disperse in the planktonic state and the final biofilm fraction is $f = 0\%$. An animated form of this Figure is shown in Movie S2. The values of the simulation parameters are given in Table S2.}\label{fig02}
\end{figure*}

\subsection*{Representative numerical simulations}

The results of a prototypical example are shown in Fig.~\ref{fig01}C and Movie S1. Consumption by the planktonic cells (green curves) rapidly establishes a steep nutrient gradient (purple) at the leading edge of the inoculum. This gradient forces the planktonic cells to then move outward via chemotaxis. In particular, they self-organize into a coherent front that expands from the initial inoculum and continually propagates, sustained by continued consumption of the surrounding nutrient---consistent with the findings of previous studies of planktonic bacteria \cite{bhattacharjee2020waves}. In this case, however, the cells also concomitantly produce autoinducer that accumulates into a growing plume (red). In some locations, the autoinducer eventually exceeds the threshold $a^*$, thus driving the formation of an immobilized biofilm (blue). Hence, at long times, $f=21\%$ of the overall population is biofilm-associated, while the remaining $1-f=79\%$ continues to disperse in the planktonic state.

Because the processes underlying motility-mediated dispersal and autoinducer-mediated biofilm formation are highly species- and environment-dependent, the values of the parameters in Eqs. \ref{eq:b1}-\ref{eq:ai} can span broad ranges---giving rise to different emergent behaviors under different conditions. Our simulations provide a way to examine how these behaviors depend on cellular concentration and motility, quantified by  $\{b_{1,0},D_{1},\chi_{1},c_{-},c_{+}\}$, nutrient availability and consumption, quantified by $\{D_{c},c_0,\kappa_1,\kappa_2,c_{\rm{char}}\}$, cellular proliferation, quantified by $\{\gamma_{1},\gamma_{2}\}$, and autoinducer production, availability, and sensing, quantified by $\{D_{a},k_1,k_2,\lambda,\alpha,\tau,a^{*}\}$. For example, implementing the same simulation as in Fig.~\ref{fig01}C, but for cells with faster nutrient consumption, yields a population that completely disperses in the planktonic state (the fraction of the population in the biofilm state at the final time of $t=20$ \si{\hour} is $f=0\%$, as shown in Fig.~\ref{fig02} and Movie S2). Conversely, when cells consume nutrient slower, a larger fraction of the population forms an immobilized biofilm ($f=52\%$, Fig. S2 and Movie S4). 

Given that the competition between motility-mediated dispersal and autoinducer-mediated biofilm formation depends sensitively on such a bewildering array of cellular and environmental factors, we ask whether these dependencies can be captured by simple, generalizable, biophysical rules. Nondimensionalization of Eqs. \ref{eq:b1}-\ref{eq:ai} yields characteristic quantities and dimensionless groups that can parameterize these dependencies, as detailed in the SI; however, given the large number of such groups, we seek an even simpler representation of the underlying processes that could unify the influence of all these different factors. To do so, we examine the fundamental processes underlying biofilm formation in our model.

\begin{figure*}
\centering
\includegraphics[width=11.4cm]{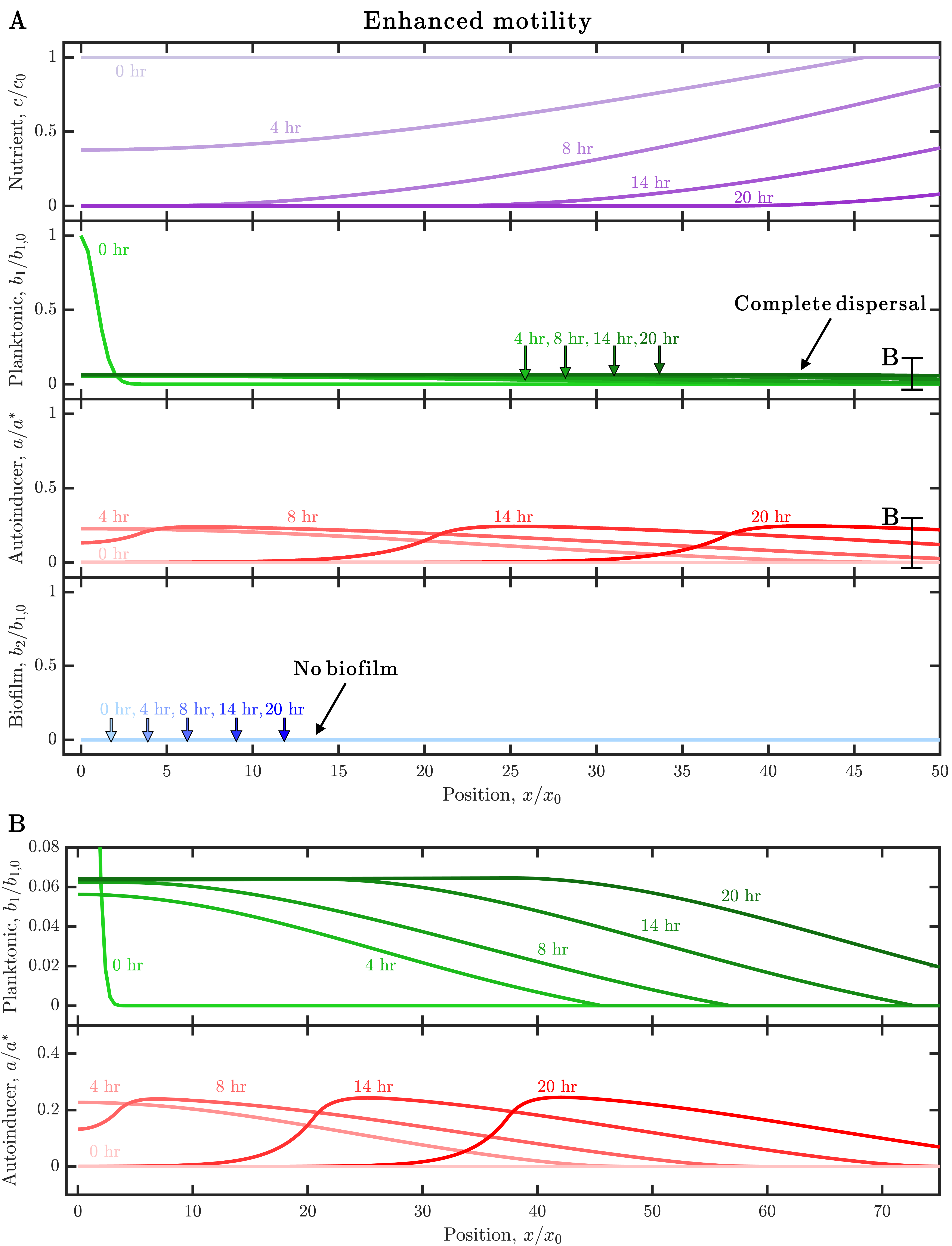}
\caption{\textbf{Enhanced motility enables cells to disperse before sufficient autoinducer accumulates, leading to complete dispersal.} \textbf{(A)} Results of the same simulation as in Fig.~\ref{fig01}C, but for faster-moving planktonic cells (larger $D_1$ and $\chi_1$). Panels and colors show the same quantities as in Fig.~\ref{fig01}C. The inoculum initially centered about the origin consumes nutrient (purple), establishing a gradient that drives outward dispersal by chemotaxis (outward moving green curves); the cells also produce autoinducer (red) concomitantly. More rapid dispersal enables the planktonic cells to ``outrun'' the growing autoinducer plume, as shown by the extended and magnified view in \textbf{(B)}. As a result, the population continues to disperse in the planktonic state and the final biofilm fraction is $f = 0\%$. An animated form of this Figure is shown in Movie S3. The values of the simulation parameters are given in Table S2.}\label{fig03}
\end{figure*}

\subsection*{Availability of nutrient for autoinducer production}

When autoinducer production is nutrient-dependent, we expect that a necessary condition for biofilm formation is that enough nutrient is available for sufficient autoinducer to be produced to eventually exceed the threshold $a^*$. To quantify this condition, we estimate two time scales: $\tau_d$, the time taken by the population of planktonic cells to deplete all the available nutrient locally, and $\tau_a$, the time at which produced autoinducer reaches the threshold for biofilm formation. While $\tau_d$ and $\tau_a$ can be directly obtained in each simulation, we seek a more generally-applicable analytical expression for both, solely using parameters that act as inputs to the model. In particular, for simplicity, we consider nutrient consumption and autoinducer production, both occurring at their maximal rates $\kappa_1$ and $k_1$ respectively, by an exponentially-growing population of planktonic cells that are uniformly-distributed in a well-mixed and fixed domain. Integrating Eqs. \ref{eq:c} and \ref{eq:ai} then yields (SI)
\begin{equation}
    \tau_d=\gamma_{1}^{-1}\ln(1+\tilde{\beta}_{1,0})
\label{nutrientdepletion}
\end{equation}
\begin{equation}
    \tau_a=\gamma_{1}^{-1}\ln\left[1-\tilde{\zeta}_{1,0}^{-1}\ln\left(1-\tilde{\eta}\right)\right].
\label{aiproduction}
\end{equation}
Three key dimensionless quantities, denoted by the tilde $(~\tilde{}~)$ notation, emerge from this calculation. The first, $\tilde{\beta}_{1,0}\equiv\gamma_1/\left(b_{1,0}\kappa_{1}/c_{0}\right)$, describes the yield of new cells produced as the population consumes nutrient---quantified by the rates of cellular proliferation and nutrient consumption, $\gamma_1$ and $b_{1,0}\kappa_{1}/c_{0}$, respectively \cite{amchin2021influence}. The second, $\tilde{\eta}\equiv \alpha a^*/k_{1}$, describes the competition between autoinducer loss and production, quantified by their respective rates $\alpha a^*$ and $k_{1}$, at the single-cell scale. The third, $\tilde{\zeta}_{1,0}\equiv\alpha b_{0}/\gamma_{1}$, describes the loss of autoinducer due to cell-surface binding as the population continues to grow, quantified by the population-scale rates of autoinducer loss and cellular proliferation, $\alpha b_{0}$ and $\gamma_{1}$, respectively; for simplicity, this quantity neglects natural degradation of autoinducer, given that the degradation rate is relatively small, with $\lambda\ll\alpha b_{0}$.

The ratio between Eqs. \ref{nutrientdepletion} and \ref{aiproduction} then defines a \textit{nutrient availability parameter}, $\tilde{\mathcal{D}}\equiv \tau_d/\tau_a$. When $\tilde{\mathcal{D}}$ is large, produced autoinducer rapidly reaches the threshold for biofilm formation before the available nutrient is depleted; by contrast, when $\tilde{\mathcal{D}}$ is small, nutrient depletion limits autoinducer production. Hence, we hypothesize that $\tilde{\mathcal{D}}\gtrsim \tilde{\mathcal{D}}^*$ specifies a necessary condition for biofilm formation, where $\tilde{\mathcal{D}}^*$ is a threshold value of order unity. The simulations shown in Figs. \ref{fig01}C, \ref{fig02}, and S2 enable us to directly test this hypothesis. Consistent with our expectation, the simulation in Fig. \ref{fig01}C is characterized by $\tilde{\mathcal{D}}=0.33$, near the expected threshold for biofilm formation; as a result, $f=21\%$. When consumption is faster as in Fig. \ref{fig02} ($\tilde{\mathcal{D}}=0.033$), the available nutrient is rapidly depleted; thus, cells disperse away before sufficient autoinducer is produced to initiate biofilm formation, and $f=0\%$. Conversely, when nutrient consumption is slow as in Fig. S2 ($\tilde{\mathcal{D}}=3.1$), nutrient continues to be available for autoinducer production, eventually driving biofilm formation, with a larger fraction $f=52\%$.

Taken together, these results support our hypothesis that $\tilde{\mathcal{D}}\gtrsim \tilde{\mathcal{D}}^*\sim1$ is a necessary condition for biofilm formation. It is not, however, a sufficient condition: repeating the simulation of Fig. \ref{fig01}C but for faster-moving cells yields a population that rapidly disperses without forming a biofilm at all ($f=0\%$, Fig. \ref{fig03}A and Movie S3)---despite having the same value of $\tilde{\mathcal{D}}=0.33$. Thus, our mathematical description of the conditions that determine biofilm formation is, as yet, incomplete.

\subsection*{Competition between motility-mediated dispersal and autoinducer accumulation}

The results shown in Fig. \ref{fig03} indicate that the ability of bacteria to move, which is not incorporated in the nutrient consumption parameter $\tilde{\mathcal{D}}$, also plays a key role in regulating whether a biofilm forms. Indeed, close inspection of Fig. \ref{fig03}A hints at another necessary condition for biofilm formation: as shown by the magnified view in Fig. \ref{fig03}B (e.g., at $t=4$ h), the leading edge of the dispersing planktonic cells extends beyond the plume of produced autoinducer. Therefore, we expect that even when sufficient nutrient is available for autoinducer production ($\tilde{\mathcal{D}}\gtrsim \tilde{\mathcal{D}}^*\sim1$), autoinducer production must be rapid enough to reach the threshold for biofilm formation before cells have dispersed away. To quantify this condition, we estimate the the time $\tau_c$ at which the motile planktonic cells begin to ``outrun'' the growing autoinducer plume. Specifically, we quantify the dynamics of the leading edge positions of the chemotactic front of planktonic cells and the autoinducer plume, $x_{1,\rm{edge}}(t)$ and $x_{a,\rm{edge}}(t)$, respectively. The front position $x_{1,\rm{edge}}(t)$ is known to depend on cellular motility, nutrient diffusion, and nutrient consumption in a non-trivial manner \cite{bergecoli,cremer2019chemotaxis,fu2018spatial,amchin2021influence}, and we are not aware of a way to compute this quantity \textit{a priori} from input parameters; instead, we extract this sole quantity from each simulation by identifying the largest value of $x$ at which $b_{1}\geq10^{-4}b_{1,0}$. While the plume position $x_{a,\rm{edge}}(t)$ can also be directly obtained in each simulation, we again develop a more generally-applicable analytical expression by assuming that the autoinducer continually diffuses from the initial inoculum: $x_{a,\rm{edge}}(t)=x_0+\sqrt{2D_{a}t}$. Then, $\tau_c$ can be directly determined as the time at which $x_{1,\rm{edge}}(t)$ begins to exceed $x_{a,\rm{edge}}(t)$.

The ratio between $\tau_c$ thereby determined and $\tau_a$, the time required for produced autoinducer to reach the threshold for biofilm formation (Eq. \ref{aiproduction}), then defines a \textit{cellular dispersal parameter}, $\tilde{\mathcal{J}}\equiv \tau_c/\tau_a$. When $\tilde{\mathcal{J}}$ is large, autoinducer accumulation is sufficiently rapid to drive biofilm formation; by contrast, when $\tilde{\mathcal{J}}$ is small, the planktonic cells rapidly disperse without forming a biofilm. Hence, we hypothesize that $\tilde{\mathcal{J}}\gtrsim \tilde{\mathcal{J}}^*$ specifies another necessary condition for biofilm formation, where $\tilde{\mathcal{J}}^*$ is, again, a threshold value of order unity. The simulations shown in Figs. \ref{fig01}C and \ref{fig03}A enable us to directly test this hypothesis. Consistent with our expectation, the simulations in Fig. \ref{fig01}C and S2 are characterized by $\tilde{\mathcal{J}}=1.6$, near the expected threshold for biofilm formation; as a result, $f>0$ in both cases. Furthermore, implementing the same simulation as Fig. \ref{fig01}C (with the same $\tilde{\mathcal{D}}=0.33$) but for slower-moving cells, characterized by a larger $\tilde{\mathcal{J}}=120$, yields a population that forms an even larger biofilm fraction $f=82\%$ (Fig. S3 and Movie S5). Conversely, when cellular dispersal is faster as in Fig. \ref{fig03}A, characterized by a smaller $\tilde{\mathcal{J}}=0.1$, the cells disperse away before sufficient autoinducer is produced to initiate biofilm formation, and $f=0\%$. Taken together, these results support our hypothesis that $\tilde{\mathcal{J}}\gtrsim \tilde{\mathcal{J}}^*\sim1$ is another necessary condition for biofilm formation.

\begin{figure*}
\includegraphics[width=17.8cm]{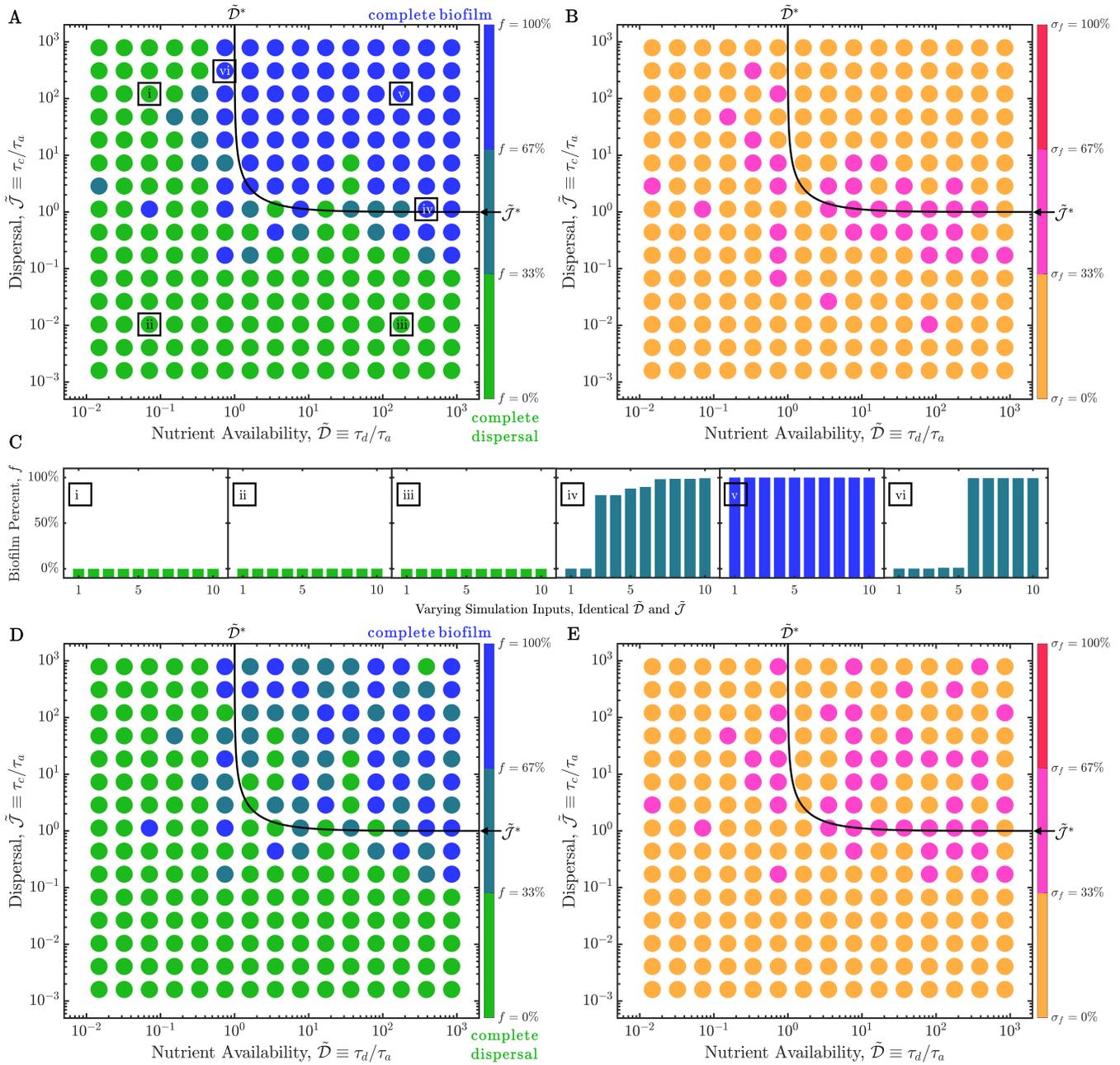}
\caption{\label{fig04} \textbf{The two states of complete dispersal by planktonic cells (green) and complete formation of a biofilm (blue) can be universally described by three dimensionless parameters.} \textbf{(A)} State diagram showing the fraction of biofilm formed, $f$, at the final time ($t=20$ \si{\hour}) for different values of the nutrient availability and cellular dispersal parameters, $\tilde{\mathcal{D}}$ and $\tilde{\mathcal{J}}$, respectively. The state diagram summarizes the results of 10,983 simulations of Eqs.~\ref{eq:b1}-\ref{eq:ai} exploring the full range of parameter values describing different bacterial species/strains and different environmental conditions (Table S1). Each point represents the mean value of $f$ obtained from multiple simulations with different parameter values, but with similar $\tilde{\mathcal{D}}$ and $\tilde{\mathcal{J}}$ (identical within each bin defined by the spacing between points). \textbf{(B)} represents the same data, but each point represents the standard deviation of the values of $f$ obtained from the same simulations. Despite the vastly differing conditions explored in each simulation, they cluster into the two states of planktonic dispersal (green) and biofilm formation (blue) when parameterized by $\tilde{\mathcal{D}}$ and $\tilde{\mathcal{J}}$. The boundary between the two states can be described by the relation $\tilde{\mathcal{D}}^*/\tilde{\mathcal{D}} + \tilde{\mathcal{J}}^*/\tilde{\mathcal{J}} \sim 1$, as shown by the black line; this relation combines the transition between the two states that occurs at both $\tilde{\mathcal{D}}^* \sim 1$ and $\tilde{\mathcal{J}}^* \sim 1$. Away from this boundary, all simulations for the same $\tilde{\mathcal{D}}$ and $\tilde{\mathcal{J}}$ collapse to have the same biofilm fraction $f$, as shown by the points in \textbf{(B)} and examples (i)-(iii) and (v) in \textbf{(C)}---confirming the universality of our parameterization.  Near the boundary, we observe some slight differences between simulations, as shown in \textbf{(B)} and examples (iv) and (vi) in \textbf{(C)}. The values of the simulation parameters for the examples in \textbf{(C)} are given in Dataset S1. The data in \textbf{(A)--(C)} correspond to a fixed value of the third dimensionless parameter $\tilde{\mathcal{S}}=50$, which describes the case of biofilm cells that produce autoinducer rapidly; repeating these simulations for the opposite case of slow autoinducer production by biofilm cells ($\tilde{\mathcal{S}}=1/50$) yields the state diagram shown in \textbf{(D)}, but for 14,351 simulation runs; again, \textbf{(E)} shows the standard deviation of the corresponding values of $f$. As shown by \textbf{(D)--(E)}, while the transition between the two states (black line) is unaffected by the change in $\tilde{\mathcal{S}}$, the transition to complete biofilm formation is more gradual. Together, the three parameters $\tilde{\mathcal{D}}$, $\tilde{\mathcal{J}}$, and $\tilde{\mathcal{S}}$ provide a full description of the onset and extent of biofilm formation across vastly different conditions.}
\end{figure*}

\subsection*{A universal biophysical threshold for biofilm formation}

Thus far, we have shown that the two conditions $\tilde{\mathcal{D}}\gtrsim \tilde{\mathcal{D}}^*$ and $\tilde{\mathcal{J}}\gtrsim \tilde{\mathcal{J}}^*$ are both necessary for biofilm formation. Is the combination of both sufficient to fully specify the conditions required for biofilm formation? To test this possibility, we implement 10,983 numerical simulations of Eqs. \ref{eq:b1}--\ref{eq:ai} exploring the full physiological ranges of the input parameters that describe cellular, nutrient, and autoinducer properties for diverse bacterial species/strains and environmental conditions (Table S1). For each simulation, we compute $\tilde{\mathcal{D}}$, $\tilde{\mathcal{J}}$, and $f$. Remarkably, despite the extensive variability in the values of the underlying parameters, all the results cluster between two states parameterized by $\tilde{\mathcal{D}}$ and $\tilde{\mathcal{J}}$, as shown in Fig. \ref{fig04}A: motility-mediated dispersal without biofilm formation ($f=0\%$, green points) when either $\tilde{\mathcal{D}}<\tilde{\mathcal{D}}^*$ or $\tilde{\mathcal{J}}<\tilde{\mathcal{J}}^*$, and biofilm formation without dispersal ($f=100\%$, blue points) when both $\tilde{\mathcal{D}}>\tilde{\mathcal{D}}^*$ and $\tilde{\mathcal{J}}>\tilde{\mathcal{J}}^*$. Many different combinations of the input parameters yield the same $(\tilde{\mathcal{D}}$, $\tilde{\mathcal{J}})$; yet, no matter the input values of these parameters, which vary over broad ranges for different cells and environmental conditions, $(\tilde{\mathcal{D}}$, $\tilde{\mathcal{J}})$ uniquely specify the resulting biofilm fraction $f$ for all points, as shown in Figs. \ref{fig04}B-C---indicating that these two dimensionless parameters reasonably encompass all the factors determining biofilm formation. We observe some exceptions at the boundary between these two states, likely because the simplifying assumptions underlying the derivation of the $\tilde{\mathcal{D}}$ and $\tilde{\mathcal{J}}$ parameters begin to break down. Nevertheless, the boundary between both states, summarized by the relation $\tilde{\mathcal{D}}^*/\tilde{\mathcal{D}}+\tilde{\mathcal{J}}^*/\tilde{\mathcal{J}}\sim1$ with $\tilde{\mathcal{D}}^*$ and $\tilde{\mathcal{J}}^*$ both $\sim1$ (black curve), thus specifies a universal biophysical threshold for biofilm formation.

\section*{Discussion}

The transition from the planktonic to biofilm state is known to depend on a large array of factors that describe cellular concentration, motility, and proliferation; nutrient availability and consumption; and autoinducer production, availability, and sensing---all of which can vary considerably for different strains/species of bacteria and environmental conditions. Therefore, quantitative prediction of the onset of biofilm formation is challenging. The biophysical model presented here provides a key step toward addressing this challenge. In particular, for the illustrative case we consider---in which cells can either disperse through active motility, retaining them in the planktonic state, or form an immobilized biofilm when exposed to sufficient autoinducer---we have shown that the onset of biofilm formation is uniquely specified by a biophysical threshold set by the two dimensionless parameters $\tilde{\mathcal{D}}$ (quantifying nutrient availability) and $\tilde{\mathcal{J}}$ (quantifying bacterial dispersal). Importantly, within the formulation of our model, this threshold is universal: many different combinations of cellular and environmental factors are described by the same $(\tilde{\mathcal{D}},\tilde{\mathcal{J}})$, and thus, yield the same onset of biofilm formation. Therefore, given a bacterial strain and set of environmental conditions, extensions of our model could help provide a way to predict whether a biofilm will form \textit{a priori}. Indeed, because the factors that define $\tilde{\mathcal{D}}$ and $\tilde{\mathcal{J}}$ can be directly measured, our work now provides quantitative principles and predictions (as summarized in Fig. \ref{fig04}) to guide future experiments.

For generality, our model also incorporates proliferation, nutrient consumption, and autoinducer production by cells after they have transitioned to the biofilm state. Hence, within our model, biofilm-produced autoinducer could also drive surrounding planktonic cells to transition to the biofilm state. In this case, we expect that the long-time fraction of the population in the biofilm state, $f$, will also depend on nutrient depletion and autoinducer production by the growing biofilm. Indeed, performing a similar calculation as that underlying the nutrient availability parameter, $\tilde{\mathcal{D}}$, yields a third dimensionless parameter, $\tilde{\mathcal{S}}\equiv\tau_{d,2}/\tau_{a,2}$; here, $\tau_{d,2}$ and $\tau_{a,2}$ describe the times at which biofilm cells have depleted all the available nutrient and produced enough autoinducer to reach the threshold for biofilm formation, respectively (SI). Thus, we hypothesize that, while the \textit{onset} of biofilm formation is specified by $(\tilde{\mathcal{D}},\tilde{\mathcal{J}})$, the final \textit{extent} of biofilm that has formed will also be described by $\tilde{\mathcal{S}}$. The results shown in Figs. \ref{fig01}--\ref{fig04}C have a fixed $\tilde{\mathcal{S}} = 50$, which describes the case of a biofilm that produces autoinducer rapidly; repeating these simulations for the opposite case of slow autoinducer production by biofilm cells, with $\tilde{\mathcal{S}} = 1/50$, yields the state diagram shown in Fig. \ref{fig04}D. In agreement with our hypothesis, while the transition to the biofilm state (black line) is not appreciably altered by the change in $\tilde{\mathcal{S}}$, the transition to complete biofilm formation ($f=1$) is more gradual in this case (compare Figs. \ref{fig04}A--B and D--E). Moreover, we note that our analysis thus far has focused on the case in which autoinducer production is nutrient-dependent; however, this process may sometimes be nutrient-independent \cite{narla2021biophysical}. In this case, we expect that our overall analysis still applies, but with the onset of biofilm formation specified by only the dispersal parameter $\tilde{\mathcal{J}}$---as confirmed in Fig. S1.

The transition from the planktonic to biofilm state is highly complex and, in many cases, has features that are unique to different species of bacteria. Nevertheless, our model provides a minimal description that can capture many of the essential features of biofilm formation more generally---thereby providing a foundation for future extensions of our work. As an illustrative example, our model considers the case in which cells produce a single autoinducer; however, some quorum sensing systems utilize multiple autoinducers \cite{miller2001quorum,miller2002parallel,pesci1997regulation}, which could be described using additional field variables and equations similar to Eq. \ref{eq:ai}. Furthermore, our model considers positive quorum sensing control in which planktonic cells irreversibly transition to the biofilm state when the local autoinducer concentration exceeds a threshold value \cite{nadell2008evolution}. However, biofilm formation is often not irreversible \cite{barraud2006involvement,kaplan2010biofilm,abdel2014bacterial}, which could be described using additional terms similar to the last terms of Eqs. \ref{eq:b1}--\ref{eq:b2}, but with the opposite sign. Similar modifications could be made to describe other species of bacteria (e.g., \textit{Vibrio cholerae}) that utilize the opposite case of negative quorum sensing control, in which biofilm cells instead transition to the planktonic state when the autoinducer accumulates above a threshold value \cite{bridges2019intragenus}.

\section*{Materials and Methods}
To numerically solve the continuum model described by Eqs. 1--4, we follow the experimentally-validated approach used in our previous work \cite{bhattacharjee2020waves,amchin2021influence}. Specifically, we use an Adams-Bashforth-Moulton predictor-corrector method in which the order of the predictor and corrector are 3 and 2, respectively. Because the predictor-corrector method requires past time points to inform future steps, the starting time points must be found with another method; we choose the Shanks starter of order 6 as described previously \cite{rodabaugh1965efficient,shanks1966solutions}. For the first and second derivatives in space, we use finite difference equations with central difference forms in rectilinear coordinates. The temporal and spatial resolution of the simulations are $\delta t=$ 0.1 \si{\second} and $\delta x=$ 20 \si{\micro\meter}, respectively; furthermore, we constrain our analysis to simulations for which the peak of the overall bacteria population moves slower than $\delta x/\delta t$. Repeating representative simulations with different spatial and temporal resolution indicates that even finer discretization does not appreciably alter the results (Fig. S4). Thus, our choice of discretization is sufficiently finely-resolved such that the results in the numerical simulations are not appreciably influenced by discretization.

To connect the simulations to our previous experiments \cite{bhattacharjee2020waves}, we choose a total extent of $1.75 \times 10^4$ \si{\micro\meter} for the size of the entire simulated system, with no-flux conditions for the field variables $b_1$, $b_2$, $c$, and $a$ applied to both boundaries at $x = 0$ and $1.75 \times 10^4$ \si{\micro\meter}. As in the experiments, we initialize each simulation with a starting inoculum of planktonic cells with a Gaussian profile defined by the maximum concentration $b_{1,0}$ at $x = 0 $ \si{\micro\meter} and a full width at half maximum of 100 \si{\micro\meter}. Nutrient is initially uniform at a fixed concentration $c_0$, and the autoinducer and biofilm concentrations are initially zero, throughout. Furthermore, following previous work \cite{amchin2021influence,dell2018growing,volfson2008biomechanical,farrell2013mechanically,klapper2002finger,head2013linear}, we also incorporate jammed growth expansion of the population in which growing cells push outward on their neighbors when the total concentration of bacteria is large enough. In particular, whenever the total concentration of bacteria (planktonic and biofilm) exceeds the jamming limit of $0.95$ cells \si{\per\micro\meter\cubed} at a location $x_i$, the excess cell concentration is removed from $x_i$ and added to the neighboring location, $x_{i}+\delta x$, where $\delta x$ represents the spatial resolution of the simulation, retaining the same ratio of planktonic to biofilm cells in the new location. We repeat this process for every location in the simulated space for each time step. 

We run each simulation for a total simulated duration of $t_{\textrm{sim}}=20$ \si{\hour}. At this final time, we use the simulation data to directly compute $f\equiv b_{2}/(b_{1}+b_{2})$, the total fraction of the population in the biofilm state. We also compute the values of the dimensionless parameters $\tilde{\mathcal{D}}$, $\tilde{\mathcal{J}}$, and $\tilde{\mathcal{S}}$ using the equations presented in the main text. We note that the autoinducer production time $\tau_a$ (Eq. \ref{aiproduction}) is only finite for $\tilde{\eta}\equiv\alpha a^*/k_1<1$; when $\tilde{\eta}\geq1$, the rate of autoinducer loss exceeds that of autoinducer production, and thus the time required to reach the threshold for biofilm formation diverges. Because both $\tilde{\mathcal{D}}$ and $\tilde{\mathcal{J}}$ are defined as $\tau_{d}/\tau_{a}$ and $\tau_{c}/\tau_{a}$, respectively, for simulations with $\tilde{\eta}\geq1$, we represent them on the state diagrams in Figs. \ref{fig04} and S1 at $(\mathcal{D},\mathcal{J})=(10^{-2},10^{-3})$, the smallest values shown on the diagrams. All of these simulations have $f=0$, as expected. Furthermore, to ensure that $t_{\textrm{sim}}$ is sufficiently long, we (i) only perform simulations with $\tau_a$ and $\tau_{a,2}$ smaller than $t_{\textrm{sim}}$, and (ii) do not include simulations with $f = 0$ but $\tau_c = \tau_{sim}$, for which sufficient time has not elapsed for planktonic cells to chemotactically disperse.

\begin{acknowledgments}
It is a pleasure to acknowledge R. K. Bay, A. M. Hancock, J. T. N. Moore, P. G-A Ott-Moore, N. A. Ott-Moore, A. Ko\v{s}mrlj, H. A. Stone, and N. S. Wingreen for stimulating discussions. This work was supported by NSF grants CBET-1941716 and EF-2124863, the Eric and Wendy Schmidt Transformative Technology Fund at Princeton, the Princeton Catalysis Initiative, a Reiner G. Stoll Undergraduate Summer Fellowship (to S.C.), and in part by funding from the Princeton Center for Complex Materials, a Materials Research Science and Engineering Center supported by NSF grant DMR-2011750. This material is also based upon work supported by the National Science Foundation Graduate Research Fellowship Program (to J.A.O.) under Grant No. DGE-1656466. Any opinions, findings, and conclusions or recommendations expressed in this material are those of the authors and do not necessarily reflect the views of the National Science Foundation. 
\end{acknowledgments}

\section*{Author contributions}
J.A.O. and S.S.D. designed the theoretical model and numerical simulations, with assistance from D.B.A. and T.B.; J.A.O. performed all numerical simulations with assistance from S.C. and D.B.A.; J.A.O., S.C., and S.S.D. analyzed the data; S.S.D. designed and supervised the overall project. All authors discussed the results and implications and wrote
the manuscript.

\section*{Competing interests}
The authors declare no competing interests. 

\setcounter{figure}{0}
\renewcommand{\figurename}{FIG.}
\renewcommand{\thefigure}{S\arabic{figure}}

\renewcommand{\tablename}{TABLE}
\renewcommand{\thetable}{S\arabic{table}}

\newpage\section*{Supplementary Information}
\subsection*{Nondimensionalizing the governing equations}
The governing equations Eqs. 1--4 are described by six variables: those describing the concentrations of planktonic bacteria ($b_1$), biofilm bacteria ($b_2$), nutrient ($c$), autoinducer molecules ($a$), as well as the one-dimensional space ($x$) and time ($t$) coordinates. Additional constants for our equations are highlighted in Table S1, with initial conditions $b_1(t=0) = b_{1,0}$, $c(t=0) = c_{0}$, and $x_{0}$ as the width of the initial planktonic inoculum. We define the dimensionless variables $\tilde{b}_1 \equiv \frac{b_1}{B_1}$, $\tilde{b}_2 \equiv \frac{b_2}{B_2}$, $\tilde{c} \equiv \frac{c}{\mathcal{C}}$, $\tilde{a} \equiv \frac{a}{A}$, $\tilde{x} \equiv \frac{x}{\mathcal{X}}$, and $\tilde{t} \equiv \frac{t}{T}$, where the tilde $(\tilde{})$ notation indicates a dimensionless quantity and the dimensional quantities $B_1$, $B_2$, $\mathcal{C}$, $A$, $\mathcal{X}$, and $T$ are not specified \textit{a priori}. Thus, in nondimensional form, Eqs. 1--4 can be represented as:
\begin{widetext}
\begin{eqnarray}
{\rm{Planktonic:}}~ \frac{\partial \tilde{b}_1}{\partial \tilde{t}} &=& \frac{D_{1}}{\left(\mathcal{X}^2/T\right)} \tilde{\nabla}^2 \tilde{b}_1 - \frac{\chi_1}{\left(\mathcal{X}^2/T\right)} \tilde{\nabla} \cdot \left(\tilde{b}_1 \tilde{\nabla} \log\left({\frac{1 + \tilde{c}/\tilde{c}_-}{1 + \tilde{c}/\tilde{c}_+}}\right)  \right) \nonumber
\\
&&+ \tilde{b}_1 (\gamma_{1}T) g(\tilde{c})- \tilde{b}_1 ({\tau}^{-1}T)\mathcal{H}\left(\tilde{a} - a^{*}/A\right)
\\
{\rm{Biofilm:}}~\frac{\partial \tilde{b}_2}{\partial \tilde{t}} &=&  (\gamma_{2}T)\tilde{b}_2  g(\tilde{c}) + (B_1/B_2)  ({\tau}^{-1}T)\tilde{b}_1\mathcal{H}\left(\tilde{a} - a^{*}/A\right)
\\
{\rm{Nutrient:}}~\frac{\partial \tilde{c}}{\partial \tilde{t}} &=& \frac{D_{c}}{\left(\mathcal{X}^2/T\right)} \tilde{\nabla}^2\tilde{c} - \left((B_1/\mathcal{C})\kappa_1 T \tilde{b}_1 + (B_2/\mathcal{C})\kappa_2 T \tilde{b}_2\right) g(\tilde{c}) 
\\
{\rm{Autoinducer:}}~ \frac{\partial \tilde{a}}{\partial \tilde{t}} &=& \frac{D_{a}}{\left(\mathcal{X}^2/T\right)} \tilde{\nabla}^2 \tilde{a} + 
\left((B_1/A)k_1 T \tilde{b}_1 + (B_2/A)k_2 T \tilde{b}_2\right) g(\tilde{c}) - T \left(\lambda + \alpha B_1 \tilde{b}_1 \right)\tilde{a}
\end{eqnarray}
\end{widetext}
where $g(\tilde{c})\equiv\frac{\tilde{c}}{{\tilde{c} + \tilde{c}_{\rm{char}}}}$. Given that the characteristic autoinducer concentration $a^*$ arises in the argument of the Heaviside step function in Eqs. S1--S2, we choose $A = a^*$. Moreover, given that the planktonic cells have a characteristic concentration $b_{1,0}$ defined by the initial inoculum, we choose $B_1 = b_{1,0}$. The fraction of the population in the biofilm state is defined as $f = {b_2}/\left(b_2 + b_1\right)$; thus, to ensure that $\tilde{f}=f$ for simplicity, we also choose $B_2 = B_1 = b_{1,0}$. Finally, given that the nutrient has a characteristic concentration $c_0$ defined by the initial saturation, we choose $\mathcal{C} = c_0$. With these choices of characteristic quantities, multiple length and time scales emerge as possible choices for $\mathcal{X}$ and $T$, respectively:
\begin{itemize}
\item Length scale: $\sqrt{T D_{1}}$, $\sqrt{T \chi_{1}}$, $\sqrt{T D_{c}}$, $\sqrt{T D_{a}}$;
\item Time scale: $\mathcal{X}^2/D_{1}$, $\mathcal{X}^2/chi_{1}$, $\mathcal{X}^2/D_{c}$, $\mathcal{X}^2/D_{a}$, $\gamma_{1}^{-1}$, $\gamma_{2}^{-1}$, $\tau$, $\frac{c_0}{b_{1,0} \kappa_1}$, $\frac{c_0}{b_{1,0} \kappa_2}$, $\frac{a^*}{b_{1,0} k_1}$, $\frac{a^*}{b_{1,0} k_2}$, $\lambda^{-1}$, $(\alpha b_{1,0})^{-1}$.
\end{itemize}
Each such choice will lead to the emergence of many different dimensionless groups characterizing this problem. Nevertheless, all these different groupings are accounted for in the dimensionless parameters $\tilde{\mathcal{D}}$, $\tilde{\mathcal{J}}$, and $\tilde{\mathcal{S}}$ described in the main text, with the exception of quantities involving the nutrient diffusivity $D_c$, planktonic-to-biofilm transition rate $\tau^{-1}$, and the natural autoinducer degradation rate $\lambda$, which have corresponding time scales that are much smaller than the other time scales of the systems considered here and are neglected from our analysis for simplicity. 

\subsection*{Derivation of the Dimensionless Parameters $\tilde{\mathcal{D}}$ and $\tilde{\mathcal{S}}$}

We first estimate the time $\tau_d$ taken for cells to deplete available nutrient through consumption. To do so, for simplicity, we consider a population of planktonic cells exponentially growing at the maximal rate $\gamma_1$, uniformly distributed in a well-mixed and fixed domain (ie., neglecting motility-mediated spreading), and consuming nutrient at the maximal rate $\kappa_1$. Thus, $\frac{dc}{dt} = - \kappa_1 b_{1,0} e^{t \gamma_1}$; integrating this equation from $t=0$ (with $c=c_0$) to $t=\tau_d$ (with $c=0$) yields Eq. 5 of the main text. 

We use a similar approach to estimate the time $\tau_a$ taken for produced autoinducer to reach the threshold for biofilm formation $a^*$. In particular, we consider the same population of planktonic cells secreting autoinducer at the maximal rate $k_1$. We neglect natural degradation of autoinducer, given that the degradation rate is relatively small compared to binding to the cell surface receptors with a second-order rate constant $\alpha$, ie., $\lambda\ll\alpha b_{0}$. The rate of autoinducer production and loss are then given by $b_{1,0} e^{t \gamma_1}\times k_1$ and $b_{1,0} e^{t \gamma_1}\times \alpha a$, respectively, ultimately yielding $\frac{d a}{d t} = b_{1,0} e^{t \gamma_1}(k_1 - \alpha a)$. Integrating this equation from $t=0$ (with $a=0$) to $t=\tau_a$ (with $a=a^*$) then yields Eq. 6 of the main text. Notably, this analytical solution for the time scale $\tau_a$ is only defined for $\tilde{\eta}\equiv\alpha a^*/k_1<1$; when $\tilde{\eta}\geq1$, the rate of autoinducer loss exceeds that of autoinducer production and secretion, and thus the time required to reach the threshold for biofilm formation diverges. Finally, the ratio of $\tau_d$ and $\tau_a$ thus derived yields the nutrient availability parameter $\tilde{\mathcal{D}}$ as described in the main text.

Thus far, we have only considered nutrient consumption by planktonic bacteria. However, cellular proliferation, autoinducer production, and nutrient consumption can also occur for cells after they have transitioned to the biofilm state, causing biofilm-produced autoinducer to also drive surrounding planktonic cells to transition to the biofilm state. Hence, we repeat the same calculations for $\tau_a$ and $\tau_d$ as described above, but now for a population of cells in the biofilm state (still with the characteristic concentration $b_{1,0}$ defined in our model), exponentially growing at the maximal rate $\gamma_2$ and consuming nutrient at the maximal rate $\kappa_2$. In this case, $\frac{dc}{dt} = - \kappa_2 b_{1,0} e^{t \gamma_2}$, and integrating this equation from $t=0$ (with $c=c_0$) to $t=\tau_{d,2}$ (with $c=0$) yields $\tau_{d,2}=\gamma_{2}^{-1}\ln(1+\tilde{\beta}_{2,0})$, where $\tilde{\beta}_{2,0}\equiv\gamma_2/\left(b_{1,0}\kappa_{2}/c_{0}\right)$ describes the yield of new biofilm cells produced as the population consumes nutrient. For the calculation of autoinducer production, we adopt a similar approach as that described above to calculate $\tau_a$, but now assuming that the biofilm surface receptors are saturated (ie., neglecting autoinducer loss). As a result, $\frac{d a}{d t} = b_{1,0} k_{2} e^{t \gamma_2}$. Integrating this equation from $t=0$ (with $a=0$) to $t=\tau_{a,2}$ (with $a=a^*$) finally yields $\tau_{a,2}=\gamma_{2}^{-1}\ln\left(1+\tilde{\theta}_{2,0}\right)$, where $\tilde{\theta}_{2,0}\equiv\frac{\gamma_2}{b_{1,0}k_{2}/a^*}$. The ratio of $\tau_{d,2}$ and $\tau_{a,2}$ thus derived then yields the parameter $\tilde{\mathcal{S}}$ as described in the main text.

\newpage\begin{figure*}
\centering
\includegraphics[width=\textwidth]{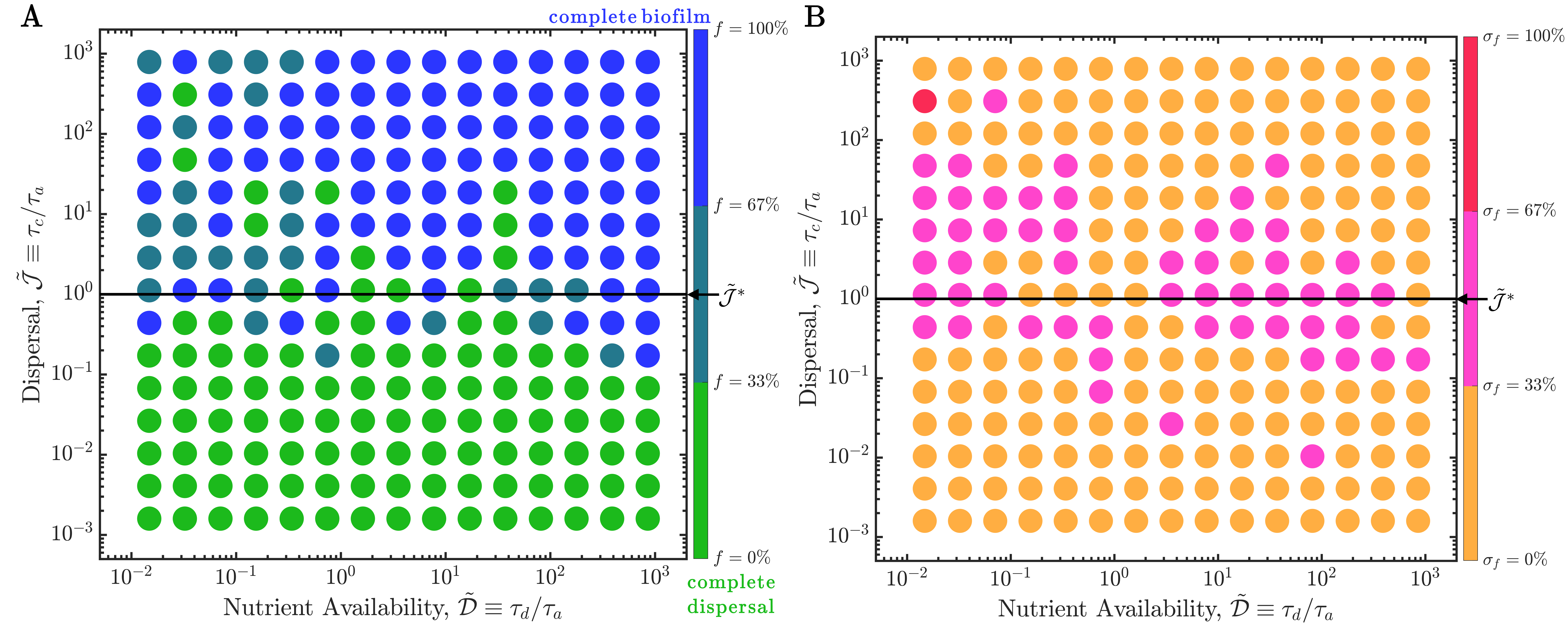}
\caption{\label{fig_si_ce}
\textbf{In the case of ``protected'' nutrient-independent autoinducer production, the transition from planktonic to biofilm states occurs at $\tilde{\mathcal{J}}\sim1$, independent of $\tilde{\mathcal{D}}$.} The model described in the main text considers autoinducer secretion to be nutrient-dependent; here, we consider the alternate case of ``protected'' nutrient-independent production at the maximal rate $k_{1}$ per cell. In this case, we do not expect the transition from the planktonic to biofilm state to depend on nutrient availability, as quantified by $\tilde{\mathcal{D}}$. Instead, the analysis presented in the main text suggests that the dispersal parameter, $\tilde{\mathcal{J}}$, alone should be sufficient to capture this transition. To test this expectation, we perform the same numerical simulations as presented in the main text, but without the Michaelis-Menten function $g(c)$ multiplying the autoinducer production term in Eq. 4. The resulting state diagrams showing the final ($t = 20$ \si{\hour}) fraction of biofilm formed, $f$, and the standard deviation of the $f$ values, as a function of $\tilde{\mathcal{D}}$ and $\tilde{\mathcal{J}}$ are shown in \textbf{(A-B)}. These diagrams summarize 8,709 separate simulations. The results confirm our expectation: in contrast to the results shown in Fig. 4, the transition does not appreciably depend on $\tilde{\mathcal{D}}$, but instead occurs near $\tilde{\mathcal{J}} \sim 1$ (black line). Thus, the theoretical framework developed in this work is more general and can be applied to bacteria with nutrient-dependent or nutrient-independent autoinducer production. }
\end{figure*}

\newpage\begin{figure*}
\centering
\includegraphics[width=11.4cm]{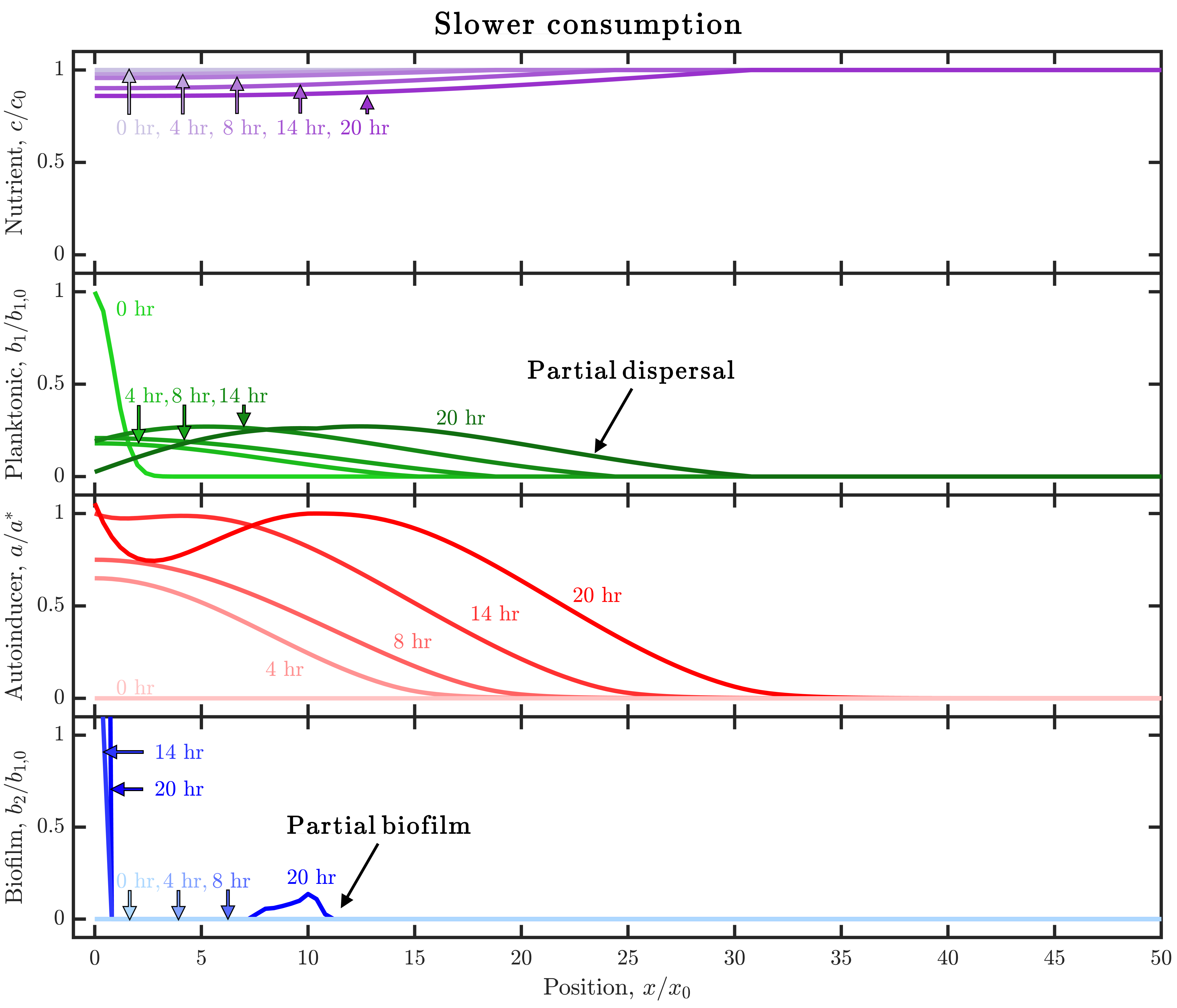}
\caption{\label{fig_si_dec_kappa} \textbf{Slower nutrient consumption allows greater autoinducer production, leading to more biofilm formation.} Results of the same simulation as in Fig. 1C, but for planktonic cells with slower nutrient consumption (smaller $\kappa_1$). Panels and colors show the same quantities as in Fig. 1C. The inoculum initially centered about the origin slowly consumes nutrient (purple), establishing a slight gradient that allows partial planktonic dispersal (green curves moving outwards); the cells also produce autoinducer (red) concomitantly. Because nutrient is consumed slowly, autoinducer production is not limited, resulting in partial biofilm formation (blue). Autoinducer has sufficiently accumulated above the threshold after $t \approx 14$ \si{\hour}, which causes a population of biofilm cells to form at the origin ($x/x_0 \approx 0$). After $20$ \si{\hour}, the biofilm population continues to grow, and additionally, autoinducer concentration exceeds the threshold concentration at $x/x_0 \approx 10$. Thus, we see a second population of biofilm cells form, centered at $x/x_0 \approx 10$. The slower nutrient consumption results in a greater final biofilm fraction than in Fig. 1C---here, $f = 52\%$. An animated form of this Figure is shown in Movie S4. The values of the simulation parameters are given in Table S2.
}
\end{figure*}

\newpage\begin{figure*}
\centering
\includegraphics[width=11.4cm]{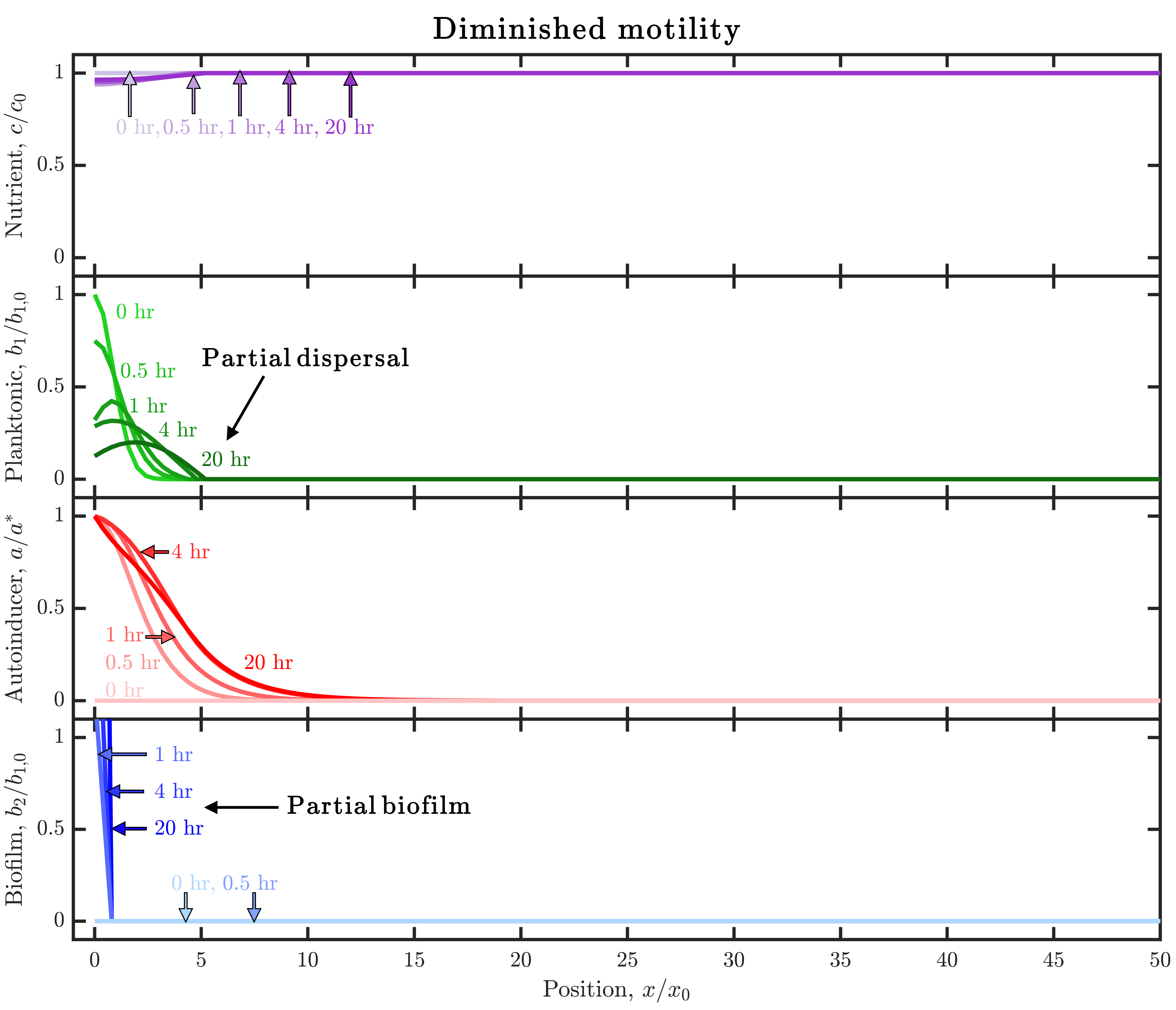}
\caption{\label{fig_si_dec_mot} \textbf{Diminished motility enables autoinducer to accumulate, resulting in increased biofilm formation.} Results of the same simulation as in Fig. 1C, but for slower-moving planktonic cells (smaller $D_1$ and $\chi_1$). Panels and colors show the same quantities as in Fig. 1C. The inoculum initially centered about the origin consumes nutrient (purple), establishing a slight gradient---however, because the motility parameters are diminished, the planktonic population (green) remains around the origin. The planktonic cells produce autoinducer (red) concomitantly, and after $1$ \si{\hour}, the autoinducer concentration exceeds the threshold concentration. Thus, some of the planktonic cells transition to biofilm cells, centered at the origin. Both the biofilm cells and planktonic cells continue to grow, produce autoinducer, and consume nutrient; the planktonic cells do not disperse due to their diminished motility, resulting in a larger fraction of biofilm cells ($f = 82\%$) than in Fig. 1C. An animated form of this Figure is shown in Movie S5. The values of the simulation parameters are given in Table S2.
}
\end{figure*}

\newpage\begin{figure*}
\centering
\includegraphics[width=\textwidth]{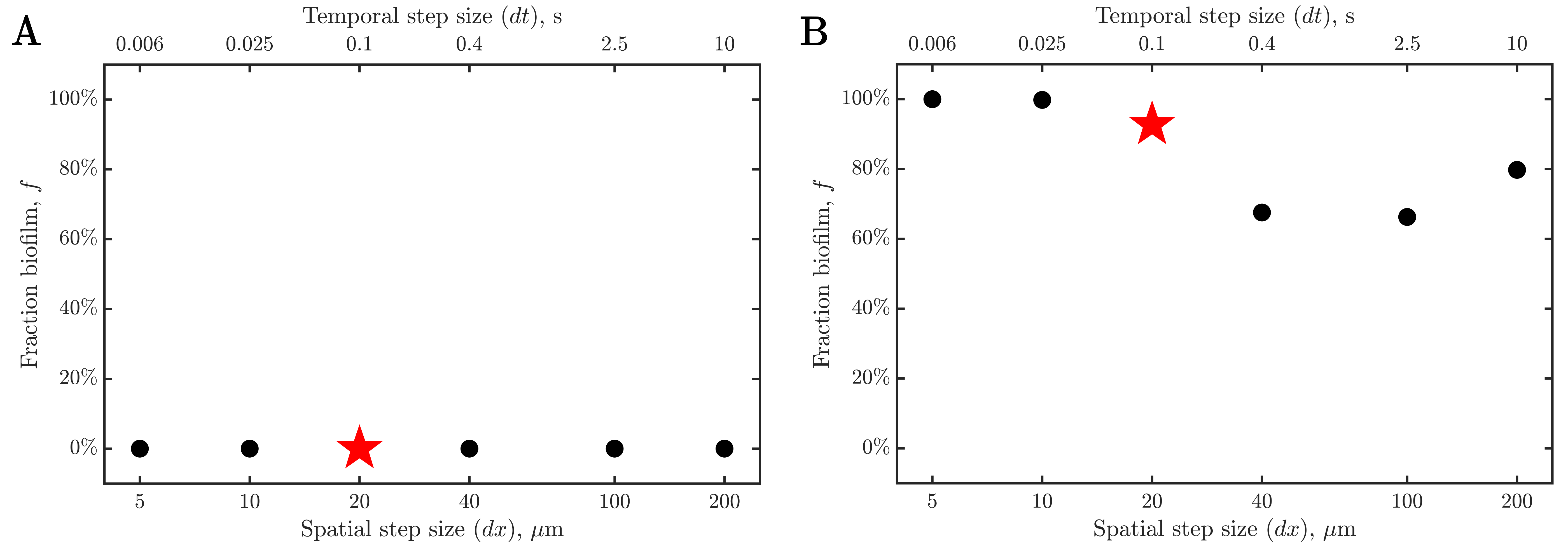}
\caption{\label{fig_si_disc_f_eq_0} \textbf{Simulation results are not appreciably influenced by our choice of discretization.} To assess the sensitivity of our results to numerical discretization, we repeat two representative simulations with \textbf{(A)} $(\tilde{\mathcal{D}},\tilde{\mathcal{J}})=(0.033,1.6)$, corresponding to $f = 0\%$, and \textbf{(B)} $(\tilde{\mathcal{D}},\tilde{\mathcal{J}})=(1.00,4.86)$, corresponding to $f = 93\%$, with varying spatial and temporal step sizes, $dx$ and $dt$, respectively. The spatial step size is changed by a factor $n$ and the corresponding temporal step size is changed by a factor $n^2$, accordingly, as indicated by the lower and upper horizontal axes, respectively. As shown in the Figure, the final biofilm fraction $f$ obtained from the simulations is not strongly sensitive to the choice of numerical discretization for spatial and temporal steps smaller than those used for the simulations presented in the main text (indicated by the red stars). Thus, our choice of discretization is sufficiently finely-resolved such that the results in the numerical simulations are not appreciably influenced by discretization.}
\end{figure*}

\newpage
\begin{table*}\centering
\caption{\label{tab:allparams} Ranges of the values of the parameters explored in our model with corresponding references.}
\begin{tabular}{lccr}
\textbf{Parameter}&
\textbf{Units}&
\textbf{Range}&
\textbf{Reference}\\
Initial bacteria concentration, $b_{1,0}$ & cells \si{\per\micro\meter\cubed} & 0.0005 - 0.94  & \cite{bhattacharjee2020waves,becker2003bacterial} \\

Maximal planktonic growth rate, $\gamma_{1}$ & \si{\per\second} &$10^{-10} - 10^{-2}$ &  \cite{bhattacharjee2020waves,wargo2008identification,wang2010monitoring}\\

Maximal biofilm growth rate, $\gamma_{2}$ &  \si{\per\second}  & $10^{-7} - 10^{-2}$& \cite{millermathmodelpbiofilm}\\

Characteristic nutrient concentration, $c_{\mathrm{char}}$ & molecules \si{\per\micro\meter\cubed} & $10^{-2} - 10^{4}$ & \cite{odell1976traveling,millermathmodelpbiofilm}\\

Initial nutrient concentration, $c_{0}$ & molecules \si{\per\micro\meter\cubed} & $10^{3} - 10^{15}$ & \cite{bhattacharjee2020waves,amchin2021influence}\\

Nutrient consumption, planktonic, $\kappa_1$ & molecules \si{\per\second} cell$^{-1}$ & $10^{0} - 10^{10}$ & \cite{odell1976traveling}\\

Nutrient consumption, biofilm, $\kappa_2$ & molecules \si{\per\second} cell$^{-1}$ & $10^{1} - 10^{9}$ & \cite{millermathmodelpbiofilm}\\

Planktonic diffusivity, $D_{1}$  & \si{\micro\meter\squared\per\second} & $0.01 - 50$ & \cite{bhattacharjee2019bacterial,berg2018random,kim1990diffusion,chen2011motility,pedit2002quantitative,olson2004quantification} \\

Planktonic chemotaxis, $\chi_{1}$  & \si{\micro\meter\squared\per\second} & $0.01 - 500$ &  \cite{amchin2021influence,pedit2002quantitative,olson2004quantification} \\

Autoinducer production rate, planktonic, $k_1$ & molecules \si{\per\second} cell$^{-1}$ & $10^{-9} - 10^{4}$ &  \cite{koerber2002mathematical}\\

Autoinducer production rate, biofilm, $k_2$ & molecules \si{\per\second} cell$^{-1}$ & $10^{-6} - 10^{8}$ & \cite{koerber2002mathematical}\\

Autoinducer threshold concentration, $a^*$ & molecules \si{\per\micro\meter\cubed} & $10^{-2} - 10^{3}$ &  \cite{pai2009optimal,weber2011noise} \\

Autoinducer diffusivity, $D_a$ & \si{\micro\meter\squared\per\second} & $10^{-2} - 10^4$ & \cite{matsui2006physical}\\

Autoinducer-receptor binding coefficient, $\alpha$ & \si{\micro\meter\cubed\per\second} cell$^{-1}$ & $10^{-7} - 10^1$ &  \cite{koerber2002mathematical}\\

Chemotaxis upper limit concentration, $c_+$ & molecules \si{\per\micro\meter\cubed} & $18000$ &  \cite{cremer2019chemotaxis,sourjik2012responding,shimizu2010modular,tu2008modeling,Kalinin2009,shoval2010fold,lazova2011response,celani2011molecular,fu2018spatial,dufour,yang2015relation,cai2016singly,chen2011motility}\\

Chemotaxis lower limit concentration, $c_-$ & molecules \si{\per\micro\meter\cubed} & $600$ &  \cite{cremer2019chemotaxis,sourjik2012responding,shimizu2010modular,tu2008modeling,Kalinin2009,shoval2010fold,lazova2011response,celani2011molecular,fu2018spatial,dufour,yang2015relation,cai2016singly,chen2011motility} \\

Autoinducer degradation rate, $\lambda$ & \si{\per\second} & $4 \times 10^{-4}$& \cite{koerber2002mathematical}\\

Planktonic transition rate, $\tau^{-1}$ & \si{\per\second} & $0.02$ & \cite{pai2009optimal,shamir2016snapshot}\\

Nutrient diffusivity, $D_c$ & \si{\micro\meter\squared\per\second} & $800$ & \cite{bhattacharjee2020waves,amchin2021influence} \\
\end{tabular}
\end{table*}

\newpage\begin{table*}\centering
\caption{\label{tab:figparams} Specific parameter values for the simulations shown in Figs. 1--3 and S2--S3.}
\begin{tabular}{lcccccr}
\textbf{Parameter}&
\textbf{Units}&
\textbf{Fig. 1C}&
\textbf{Fig. 2}&
\textbf{Fig. 3}&
\textbf{Fig. S2}&
\textbf{Fig. S3}
\\
Initial bacteria concentration, $b_{1,0}$ &  cells \si{\per\micro\meter\cubed} & 0.0016  & 0.0016 & 0.0016 & 0.0016 & 0.0016 \\

Maximal planktonic growth rate, $\gamma_1$ & \si{\per\second} & $3.6 \times 10^{-5}$ & $3.6 \times 10^{-5}$ & $3.6 \times 10^{-5}$ & $3.6 \times 10^{-5}$ & $3.6 \times 10^{-5}$\\

Maximal biofilm growth rate, $\gamma_2$ & \si{\per\second} & $3.7 \times 10^{-6}$ & $3.7 \times 10^{-6}$ & $3.7 \times 10^{-6}$ & $3.7 \times 10^{-6}$ & $3.7 \times 10^{-6}$\\

Characteristic nutrient concentration, $c_{\mathrm{char}}$ & molecules \si{\per\micro\meter\cubed} & 0.21 & 0.21 & 0.21 & 0.21 & 0.21 \\

Initial nutrient concentration, $c_0$ & molecules \si{\per\micro\meter\cubed} & $4.3 \times 10^5$ & $4.3 \times 10^5$ & $4.3 \times 10^5$ & $4.3 \times 10^5$ & $4.3 \times 10^5$ \\

Nutrient consumption, planktonic, $\kappa_1$ & molecules \si{\per\second} cell$^{-1}$ & $1.4 \times 10^6$ & $1.4 \times 10^7$ & $1.4 \times 10^6$ & $1.4 \times 10^5$& $1.4 \times 10^6$\\

Nutrient consumption, biofilm, $\kappa_2$ & molecules \si{\per\second} cell$^{-1}$ & 820 & 820 & 820 & 820 & 820\\

Planktonic diffusivity, $D_{1}$  & \si{\micro\meter\squared\per\second} & 5 & 5 & 50 & 5 & 0.5\\

Planktonic chemotaxis, $\chi_{1}$  & \si{\micro\meter\squared\per\second} & 0.01 & 0.01 & 0.1 & 0.01& 0.001\\

Autoinducer production rate, planktonic, $k_1$ & molecules \si{\per\second} cell$^{-1}$ & 3.4 & 3.4 & 3.4 & 3.4 & 3.4\\

Autoinducer production rate, biofilm, $k_2$ & molecules \si{\per\second} cell$^{-1}$ & 0.45 & 0.45 & 0.45 & 0.45 & 0.45\\

Autoinducer threshold concentration, $a^*$ & molecules \si{\per\micro\meter\cubed}& 3.1 & 3.1 & 3.1 & 3.1 & 3.1\\

Autoinducer diffusivity, $D_a$ & \si{\micro\meter\squared\per\second}& 5.0 & 5.0 &5.0 & 5.0 & 5.0\\

Autoinducer-receptor binding coefficient, $\alpha$ & \si{\micro\meter\cubed\per\second} cell$^{-1}$ & 0.061 & 0.061 & 0.061 & 0.061 & 0.061\\

Chemotaxis upper limit concentration, $c_+$ & molecules \si{\per\micro\meter\cubed} & 18000 & 18000 & 18000 & 18000 & 18000\\

Chemotaxis lower limit concentration, $c_-$ & molecules \si{\per\micro\meter\cubed} & 600 & 600 & 600 & 600 & 600\\

Autoinducer degradation rate, $\lambda$ & \si{\per\second}& $4 \times 10^{-4}$ & $4 \times 10^{-4}$ & $4 \times 10^{-4}$ & $4 \times 10^{-4}$ & $4 \times 10^{-4}$\\

Planktonic transition rate, $\tau^{-1}$ & \si{\per\second} & 0.02 & 0.02 & 0.02 & 0.02 & 0.02\\

Nutrient diffusivity, $D_c$ & \si{\micro\meter\squared\per\second} & 800 & 800 & 800 & 800 & 800\\
\end{tabular}
\end{table*}

\newpage
\providecommand{\noopsort}[1]{}\providecommand{\singleletter}[1]{#1}%

\end{document}